\documentclass[lettersize,journal]{IEEEtran}
\usepackage{amsmath,amsfonts}
\usepackage{algorithmic}
\usepackage{algorithm}
\usepackage{array}
\usepackage[caption=false,font=normalsize,labelfont=sf,textfont=sf]{subfig}
\usepackage{textcomp}
\usepackage{stfloats}
\usepackage{url}
\usepackage{verbatim}
\usepackage{graphicx}
\usepackage{cite}
\usepackage{multirow}
\usepackage{colortbl}
\usepackage[table,xcdraw]{xcolor} 
\usepackage{color}
\usepackage{booktabs}
\begin{document}

\title{Human-Inspired Computing for Robust and Efficient Audio-Visual Speech Recognition}

\author{Qianhui Liu,
    Jiadong Wang,
    Yang Wang,
    Xin Yang,
    Gang Pan, ~\IEEEmembership{Senior Member,~IEEE,}
    Haizhou Li, ~\IEEEmembership{Fellow,~IEEE,}
\thanks{
}
}

\maketitle

\begin{abstract}
Humans excel at audiovisual speech recognition (AVSR), motivating the development of human-inspired computing for robust and efficient AVSR models. Spiking neural networks (SNNs),  mimicking the brain's information-processing mechanisms, offer a promising foundation.
However, research on SNN-based AVSR remains limited, with most audio-visual methods focusing on object or digit recognition. These methods oversimplify multimodal fusion, neglecting modality-specific characteristics and interactions. Additionally, they often rely on future information, increasing recognition latency and limiting real-time applicability. Inspired by human speech perception, this paper proposes a novel human-inspired SNN named HI-AVSNN for AVSR, incorporating three computing characteristics: spike activity, cueing interaction, and causal processing. For cueing interaction, we introduce a Spike-Driven Visual-Cued Speech Processing (sVCSP) scheme, where visual features hierarchically guide speech processing to enhance critical features. For causal processing, we align the temporal dimensions of SNN with audio-visual inputs and apply temporal masking to ensure only past and current information is used. For spike activity, in addition to SNNs, we incorporate event cameras to capture lip movements as spikes, efficiently encoding visual data like the human retina. Experiments on two event-based AVSR datasets demonstrate our method outperforms existing audio-visual SNN fusion techniques, showcasing the effectiveness, robustness, and efficiency achieved through our human-inspired computing.
\end{abstract}


\section{Introduction}
\label{sec:intro}

\IEEEPARstart{H}{uman} intelligence, refined over millennia, has evolved into an extraordinarily efficient system capable of solving various tasks. One fundamental task essential for daily communication is audiovisual speech recognition (AVSR), where the brain effectively integrates speech and visual information to enhance speech perception, especially in noisy environments \cite{wang2024predict,yang2024privacy}. This remarkable ability inspires us to explore new computing paradigms that replicate the human brain's mechanisms, aiming to develop more robust and efficient AVSR models.

Spiking neural networks (SNNs), regarded as the third generation of neural networks, mimic the brain’s information communication mechanisms \cite{roy2019towards,stauffer2022spikebase}. Unlike traditional artificial neural networks (ANNs) based on continuous floating-point values, SNNs transmit information between neurons using discrete signal timing, known as spikes. This event-driven and inherently sparse approach boosts energy efficiency. Furthermore, with their intrinsic temporal characteristics, SNNs excel at processing spatio-temporal information, making them particularly advantageous for speech recognition applications \cite{liu2022event,aung2023deepfire2}. These capabilities make SNNs promising models for emulating the human brain's remarkable AVSR abilities.

However, research on SNNs for AVSR remains limited. Existing audio-visual multimodal SNNs \cite{liu2022event,yu2022multimodal,jiang2023cmci,guo2023transformer} primarily focus on object or digit recognition, and their application to AVSR presents several challenges. First, most existing methods simply concatenate or add features from both modalities. They treat all features equally, overlooking the unique characteristics and interactions of the auditory and visual modalities. Second, some studies rely on using future information for current recognition. This necessitates the system to wait for all data to be input before processing can begin, preventing immediate recognition and increasing latency. Even the only existing SNN-based AVSR method \cite{yu2022multimodal} suffers from these two weaknesses. As a result, despite using spikes, existing audio-visual multimodal SNNs do not fully explore and mimic how the human brain processes the audio and visual modalities.

\begin{figure*}
    \centering
    \includegraphics[width=0.85\linewidth]{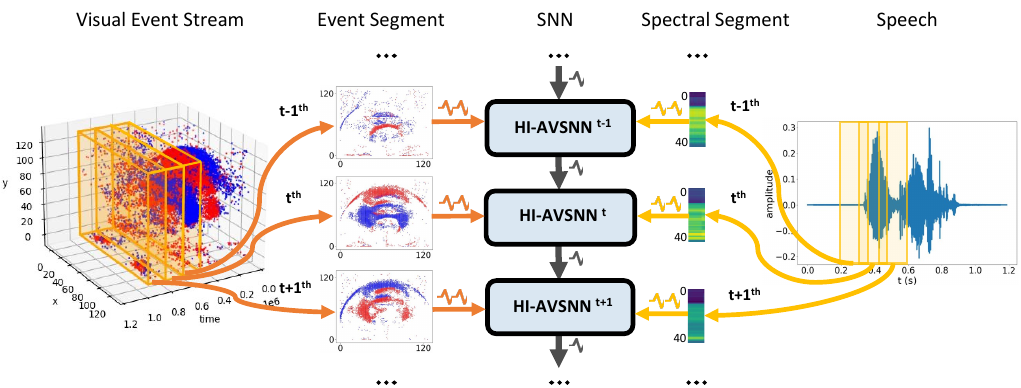}
    \caption{Temporal alignment of the SNN's processing with visual and speech inputs, enabling causal processing. The visual event stream is divided into multiple segments, with each event segment corresponding to one timestep and fed into the HI-AVSNN at that timestep. Similarly, each spectral segment from speech is input to the SNN at its corresponding timestep. At timestep $t$, the SNN processes only the $t$-th event and spectral segment inputs, along with the SNN state from the previous timestep. This alignment ensures that the system uses only past and current information, enabling real-time processing and reducing latency.}
    \label{fig:temporal}
\end{figure*}

Studies on human speech perception reveal the possible function of visual modality in enhancing speech processing. When one speaks, lip movements are found to precede the arrival of the voice \cite{varghese2012visual,golumbic2013visual}. This enables listeners to form a preliminary understanding from lip movements before hearing, thus narrowing potential word candidates \cite{wang2024predict}. Such findings led us to recognize the distinct roles of visual and auditory modalities, challenging previous approaches that relied on simple concatenation and addition. Rather than providing distinct features for direct integration with speech, visual input serves as anticipatory cues, guiding auditory attention toward the most relevant speech components. This also aligns with the previous findings \cite{schwartz2004seeing}.
Furthermore, human brain processes speech in real time, without waiting for speakers to finish speaking before beginning to process. Human short-term memory \cite{jonides2008mind} enables individuals to integrate current and past inputs for decision-making, thereby enhancing real-time speech comprehension capabilities. By leveraging these insights, we will make steps toward the human-inspired computing paradigm for advanced AVSR systems.

In this paper, we propose a novel human-inspired spiking neural network named HI-AVSNN for audio-visual speech recognition. HI-AVSNN incorporates three key computing characteristics that simulate human brain speech perception: spike activity, cueing interaction and causal processing.
Firstly, to facilitate effective interaction between visual and auditory modalities, we propose a Spike-Driven Visual-Cued Speech Processing (sVCSP) scheme. This scheme enables spiking visual features to dynamically guide speech processing, directing attention to the most relevant information. By hierarchically providing visual cues to interact with speech features, our HI-AVSNN selectively highlights key speech features, thereby improving both accuracy and robustness. Secondly, to ensure causal processing, HI-AVSNN aligns the timesteps of SNN with the time dimension of the visual and speech features, as shown in Fig. \ref{fig:temporal}. This alignment naturally restricts the system to utilizing only past and current information. We additionally utilize temporal masking to ensure that attention generated by sVCSP adheres to this temporal constraint. Finally, for spike-based computation, in addition to using SNNs, we incorporate the event camera to capture lip movement as spikes. Event cameras detect brightness changes and output spikes asynchronously, emulating the function of the human retina \cite{gallego2020event}. This allows the system to focus on lip movements while ignoring static backgrounds, leading to more precise and efficient visual acquisition for AVSR. 

We evaluate our proposed HI-AVSNN on two event-based audiovisual speech recognition datasets, DVSlip-Audio and EventLRW, which are both introduced here for the first time in AVSR. Experimental results demonstrate
that our proposed human-inspired fusion method outperforms existing audio-visual SNN fusion techniques, showcasing the effectiveness, robustness, and efficiency achieved through our human-inspired computing paradigm.

We organize the rest of this paper as follows. Section \ref{sec:related work} reviews the related work on lip reading, speech recognition, and audio-visual recognition. Section \ref{sec:method} details our HI-AVSNN framework and how it effectively embodies the three computing characteristics. Section \ref{sec:exp} presents comprehensive experimental results, and Section \ref{sec:conclusion} summarizes our conclusions.

\section{Related Work}
\label{sec:related work}
In this section, we first investigate the existing event-based lip-reading and speech recognition computing paradigms respectively. Then we introduce the computing paradigm for SNN-based audio-visual multimodal recognition.
 \subsection {Event-based Computing Paradigm for Lip Reading}
 The visual inputs for the proposed HI-AVSNN are from the event camera. Event cameras record pixel-level brightness changes on a logarithmic scale and output asynchronous events (spikes), inspired by the mechanism of the human retina \cite{gallego2020event}. Unlike conventional cameras, event cameras respond only to motion, significantly reducing memory usage and energy consumption \cite{kirkland2022unsupervised}. Additionally, the high dynamic range of event cameras (e.g., 120dB for DAVIS346) allows for reliable visual capture under extreme lighting conditions \cite{gallego2020event,yang2023learning}. Due to these advantages, event cameras have been explored for lip reading in recent studies. Tan \textit{et al.} \cite{tan2022multi} first studied the event-based lip reading and introduced a multi-grained spatio-temporal feature perceived network, demonstrating the benefits of event cameras for lip reading tasks. Given their event-driven processing and biological plausibility, SNNs are naturally well-suited to integrate with event cameras. Building on this, Bulzomi \textit{et al.} \cite{bulzomi2023end} recently proposed the first SNN designed for event-based lip-reading using a similar architecture as \cite{tan2022multi}. Liu \textit{et al.} \cite{liu2024intelligent} further proposed a spatial-temporal attention scheme integrated with triplet loss to enhance the ability to distinguish visually similar words.

\begin{figure*}
    \centering
    \includegraphics[width=1\linewidth]{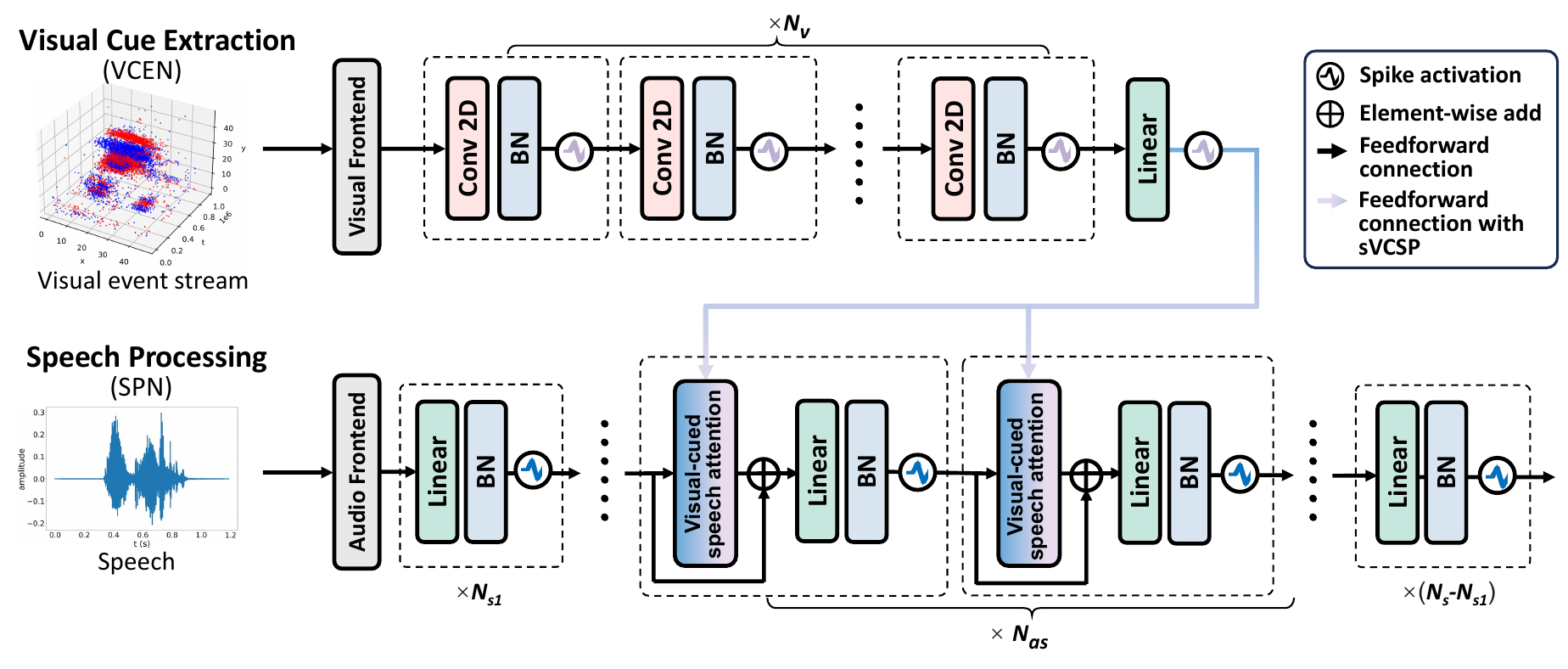}
    \caption{The proposed HI-AVSNN architecture, which consists of a visual cue extraction subnet (VCEN) and a speech processing subnet (SPN), integrated through the proposed Spike-Driven Visual-Cued Speech Processing (sVCSP) scheme. $N_v$, $N_{as}$ and $N_s$ represent the number of visual blocks, attention speech blocks and speech blocks, respectively. }
    \label{fig:framework}
\end{figure*}

\subsection{Computing Paradigm for Speech Recognition}
Automatic speech recognition (ASR) has advanced significantly with improvements in signal processing and neural network methods \cite{kheddar2024automatic}. Raw speech undergoes feature extraction using techniques like Mel-frequency cepstral coefficients (MFCC) \cite{li2024audio} and filter banks (Fbank) \cite{bittar2022surrogate} to generate spectral features. Convolutional neural networks (CNNs) \cite{pellegrini2023adapting,ZhuWCZ022}, recurrent neural networks (RNNs) \cite{swietojanski2023variable,yang2021decentralizing}, and long short-term memory (LSTM) networks \cite{cui2021federated, tsai2019using} have been explored used to process the temporal dynamics of speech information. More recently, SNNs have emerged as a compelling approach for processing speech. Some studies have shown promising results using ANN-to-SNN conversion algorithms \cite{wu2020deep, yilmaz2020deep, yang2022deep}. However, these methods do not fully utilize the temporal strengths of SNNs, as they rely on approximating the activation patterns of ANNs rather than leveraging the unique spiking dynamics of SNNs. In contrast, Bittar \textit{et al.} \cite{bittar2022surrogate} proposed recurrent spiking neurons that can use directly-trained SNN to achieve performance comparable to state-of-the-art ANNs, highlighting the potential of SNNs in ASR. 

\subsection{SNN-based Computing Paradigm for Audio-Visual Multimodal Recognition}
Zhang \textit{et al.} \cite{zhang2020efficient} proposed an audio-visual digit recognition method that first trains the SNNs for unimodal and then integrates the two modalities through excitatory and inhibitory lateral connections. Liu \textit{et al.} \cite{liu2022event} introduced an event-based multimodal SNN that concatenates two modality features and utilizes an attention mechanism to dynamically allocate the weights to two modalities. Jiang \textit{et al.} \cite{jiang2023cmci} proposed a cross-modality current integration for the multimodal SNN, which adds the currents from two modalities to fuse the information. However, these networks are simple in structure and limited to digit recognition tasks. Guo \textit{et al.} \cite{guo2023transformer} proposed a multimodal object recognition SNN, which employs audio-visual and visual-audio cross-attention before concatenation to synchronize two modalities. Yu \textit{et al.} \cite{yu2022multimodal} proposed the first SNN-based AVSR that integrates two modalities using concatenation, followed by sigmoid-based attention applied to the concatenated features. Both of these two works rely on concatenation operations and future information, overlooking the unique characteristic of each modality and introducing latency as the model needs to wait for the entire input sequence before processing can begin. 

Inspired by human speech perception, we propose a new paradigm for processing AVSR using SNNs. In our HI-AVSNN, the fusion of information from different modalities is not based on concatenation; instead, visual information provides cues to auditory processing, guiding which features should receive focused attention. Additionally, our HI-AVSNN employs causal processing, utilizing only past and current information. By more closely aligning with the natural paradigm of human speech processing, our method holds greater potential for advancing AVSR systems.

\section{The Proposed HI-AVSNN}
\label{sec:method}

 As illustrated in Fig. \ref{fig:framework}, HI-AVSNN consists of a visual cue extraction subnet (VCEN) and a speech processing subnet (SPN), integrated using a proposed Spike-Driven Visual-Cued Speech Processing (sVCSP) scheme. The VCEN takes the visual events triggered by lip movements and generates embeddings as visual cues. The SPN receives the speech input and processes the speech features in conjunction with the visual cues for recognition. The sVCSP fuses the visual and speech information, hierarchically enhancing the speech features by leveraging visual cues. The following subsections will detail these components and elaborate on how HI-AVSNN effectively embodies the three computing characteristics of human speech perception. 

\subsection{Visual Cue Extraction}
To generate the visual cues that aid speech recognition, we first employ a visual frontend to segment the event streams triggered by lip movements into $T$ timesteps and augment the data following the approach of  \cite{tan2022multi}. We then adopt $N_v$ visual blocks to extract visual features. Each visual block consists of convolution and batch normalization layers with spiking neurons. Various spiking neuron models have been developed, ranging from the simplest Integrate-and-Fire (IF) \cite{yu2023ttfs} to the more sophisticated Hodgkin–Huxley (H-H) model \cite{izhikevich2003simple}. In this paper, we choose the simple and widely used spiking Leaky Integrate-and-Fire (LIF) model \cite{wu2018spatio}, whose dynamics can be defined as:
\begin{equation}
  {u}^{n, t + 1} = \tau {u}^{n,t} + {W}^n {x}^{n-1, t+1}
\end{equation}
\begin{equation}\label{fneuron2}
	x^{n,t + 1} = \Theta(u^{n, t + 1}-V_{th})
\end{equation}
\begin{equation}
    {u}^{n, t + 1} = {u}^{n, t + 1} (1 - {x}^{n,t + 1})
\end{equation}
where ${u}^{n,t}$ denotes the membrane potential at time $t$ for the neurons in the $n$-th visual block, ${x}^{n-1,t}$ represents the input, ${W}^n$ denotes the synaptic weights, $\tau$ is the decay constant, and $\Theta$ is the Heaviside step function. When the membrane potential ${u}^{n, t + 1}$ exceeds the threshold $V_{th}$, the corresponding neurons emit a spike, after which their membrane potential is reset to $0$. The resulting spike ${x}^{n,t + 1}$ will be used as the input to the next layer. At the end of VCN, we employ a fully-connected (FC) layer to generate the visual embeddings $\phi \in \mathbb{R}^{T \times C}$, where $C$ represents the embedding dimensions, corresponding to the number of classes to be recognized. This embedding carries information about the visual aspects of the speech, which will be used as the visual cues for speech recognition.

\subsection{Speech Processing}
Audio frontend is employed to transform the raw waveforms into audio feature spikes. Given the widespread use of spectral features to represent acoustic characteristics, our frontend adopts the Filterbank (Fbank) method \cite{bittar2022surrogate} to generate spectral features. Fbank processes the speech into $T$ time segments. To capture the rich temporal characteristics of the speech, the frontend encodes the Fbank features using two layers of recurrent spiking neurons (RLIF), which possess enhanced dynamic representation capabilities \cite{bittar2022surrogate,liu2022event,jiang2023cmci}. The difference between RLIF and LIF is that the Equation (1) is modified as follows:
\begin{equation}
{u}^{n, t + 1} = \tau {u}^{n,t} + {W}^n {x}^{n-1, t+1} + {V}^n {x}^{n, t}
\end{equation}
where ${V}^n$ is the recurrent weights of the $n$-th layer. 

SPN consists of two types of blocks: speech block and attention speech block. Each speech block includes a linear layer and a batch normalization layer with spiking neurons. The attention speech block extends the functionality of the speech block by incorporating a visual-cued speech attention (VCSA) module, enabling enhanced speech processing guided by visual cues. The details of the VCSA module and its role within the sVCSP scheme are provided in Section \ref{sec:sVCAP}.

\subsection{Spike-Driven Visual-Cued Speech Processing (sVCSP) Scheme}
\label{sec:sVCAP}
\begin{figure}
    \centering
\includegraphics[width=1.0\linewidth]{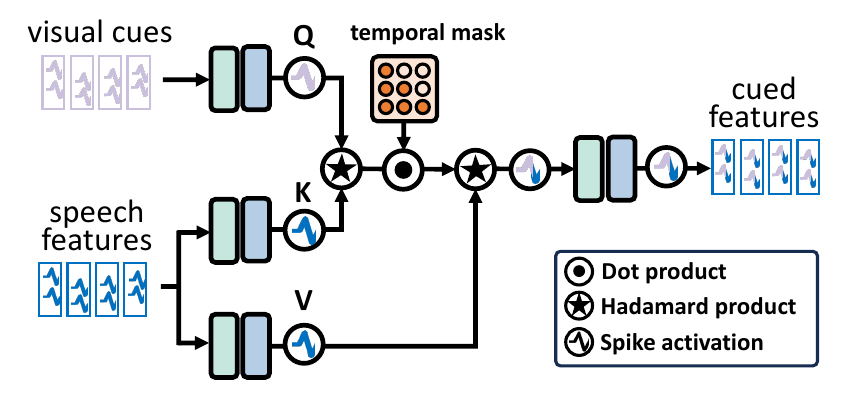}
    \caption{Visual-cued speech attention (VCSA)  module in sVCSP scheme. The visual cues, speech features, and cued features are all spike-based. The temporal mask prevents future information when computing visual-cued speech attention, ensuring causal processing.}
    \label{fig:VCAA}
\end{figure}
To emulate the human speech perception function where visual information cues listeners to focus on relevant speech signals \cite{schwartz2004seeing,wang2024predict}, using simple concatenation or addition, as done in previous works \cite{yu2022multimodal,jiang2023cmci}, is insufficient. To address this, we propose the Spike-Driven Visual-Cued Speech Processing (sVCSP) scheme, which comprises two steps: (1) fusion of visual and speech information by utilizing spiking visual cues to enhance relevant spiking speech features through the VCSA module, as shown in Fig. \ref{fig:VCAA}; and (2) refining speech focus hierarchically by applying VCSA across multiple layers, thereby improving the quality and reliability of extracted features. Mathematically, the sVCSP scheme can be expressed as follows: 
\begin{equation}
\begin{aligned}
CS^{n} &= VCSA^n(\phi, S^{n-1}), \\
n &\in \{\text{layers of attention speech blocks}\}.
\end{aligned}
\end{equation}
where $\phi$ represents the visual cues derived from VCN, $S^{n-1}$ is the input to the $n$-th attention speech block, and $CS^{n}$ denotes the cued features. The value range of $n$ corresponds to the layer number of applying attention speech block. Thus, the operation of the attention speech block can be expressed as:
\begin{equation}
\begin{aligned}
X^n &= VCSA^n(\phi, S^{n-1}) + S^{n-1} \\
S^{n} &= SN(BN(Linear(X^n)))
\label{equ:at}
\end{aligned}
\end{equation}
where BN is the batch normalization and SN is the spike activations.

Within the VCSA module of $n$-th attention speech block, the spiking visual cues $\phi \in \mathbb{R}^{T \times C}$ serve as the query, while the spiking speech features $S^n \in \mathbb{R}^{T \times L}$ are used as the key and value. This enables our VCSA to dynamically weigh the speech features based on the visual cues, enhancing the model's ability to focus on the most relevant speech information.
Specifically, query $Q$ is calculated by a learnable linear matrix $W_Q \in \mathbb{R}^{C \times D}$ with $\phi \in \mathbb{R}^{T \times C}$. Key $K$ and value $V$ are calculated by $W_K, W_V \in \mathbb{R}^{L \times D}$ with $S^n \in \mathbb{R}^{T \times L}$ respectively. This process is formulated as:
\begin{equation}
\begin{aligned}
Q^n &= SN_Q(BN(\phi W_Q^n)) \\
K^n &= SN_K(BN(S^{n-1} W_K^n)) \\
V^n &= SN_V(BN(S^{n-1} W_V^n))
\end{aligned}
\end{equation}
where $Q,K,V \in \mathbb{R}^{T \times D}$. Since both inputs $\phi$ and $S^n$ are spike tensors, the calculation here only includes addition operations. Following the \cite{zhouspikformer}, we add a scaling factor $s$ to regulate the magnitude of the matrix multiplication result, defined as:
\begin{equation}
SA'^n = SN\left((Q^n K^{n\mathrm{T}}) V^n \cdot s\right)
\end{equation}
$s$ in our work can be trained to better control the SA$'$ result.
Notably, softmax is removed from spike attention \cite{zhouspikformer}, as the attention map generated by $Q$ and $K$ is naturally non-negative. 

However, without constraints, attention considers all tokens in the sequence. This unrestricted attention introduces leakage of future information, which is inconsistent with causal processing. To enforce causality, we apply temporal masking to the attention weights, ensuring that each token only pays attention to itself and past tokens while ignoring future ones. The revised attention is formulated as:
\begin{equation}
SA'^n = SN\left(\text{mask} \cdot (Q^n K^{n\mathrm{T}}) V^n \cdot s\right)
\end{equation}
The mask is a lower triangular matrix:
\begin{equation}
\text{mask}_{i,j} =
\begin{cases}
1, & \text{if } j \leq i \\
0, & \text{if } j > i 
\end{cases}
\end{equation}
where $i$ and $j$ represent the row and column index of the mask matrix.
Unlike traditional attention mechanisms that require softmax normalization and thus need to assign $-\infty$ attention weights to masked positions, our binary mask directly assigns 1 to retain information and 0 to filter out future tokens. The VCSA ends with a linear feedforward:
\begin{equation}
CA^n = SN\left(BN^n(\text{Linear}(SA'^n))\right)
\end{equation}
The cued features are then added to the original input of VCSA and passed through the subsequent linear and batch normalization layers in the attention speech block, as expressed in Equation (\ref{equ:at}).

\subsection{Causal Processing and Training Strategy}

To implement causal processing, we align the timesteps of our HI-AVSNN with the temporal dimensions of the visual and speech features, as illustrated in Fig. \ref{fig:temporal}. Both the visual and speech inputs are divided into $T$ segments, with the $t^{th}$ segment of visual events and speech features serving as input for the $t^{th}$ timestep of SNN. Since the SNN processes only the input at the current timestep along with its previous state, this alignment inherently prevents the system from using future information. Furthermore, the masking mechanism in VCSA module reinforces causality by ensuring that attention weights are computed based solely on past and current data.
Existing audio-visual SNN methods \cite{guo2023transformer,yu2022multimodal}, however, process auditory features by treating them as static images, disregarding the temporal characteristics. Moreover, \cite{guo2023transformer} does not apply masking in attention, which allows the use of future information. As a result, these non-causal approaches require waiting for the entire input sequence to be available before processing, leading to increased system latency. Our design, in contrast, guarantees that HI-AVSNN processes data in a causal, real-time manner.

In HI-AVSNN, we employ the direct training strategy for SNNs to preserve their inherent temporal characteristics. The loss function for HI-AVSNN is defined as
\begin{equation}
	L =  CE\left(\frac{1}{T} \sum_{t=1}^{T} O(t), y\right),
\end{equation}
where $CE$ is the cross-entropy function, $O(t)$ is the output at $t$-th timestep from SPN and $y$ represents the target label. During training, we update the model parameters using spatial-temporal backpropagation (STBP) \cite{wu2018spatio}:
\begin{equation}
    \frac{\partial L}{\partial \textbf{W}} = \sum_{t} \frac{\partial L}{\partial \textbf{x}^t} \frac{\partial \textbf{x}^t}{\partial \textbf{u}^t} \frac{\partial \textbf{u}^t}{\partial \textbf{W}},
\end{equation}
Due to the non-differentiable nature of spike activities, the term $\frac{\partial \textbf{x}^t}{\partial \textbf{u}^t}$ does not exist. In this
work, we employ a triangular function \cite{wu2018spatio} to approximate the gradient of the spike function:
\begin{equation}
    \frac{\partial x^t}{\partial u^t} = h(u^t) = \frac{1}{\gamma^2} \max(0, \gamma - |u^t - V_{th}|)
\end{equation}
where $\gamma$ is the constraint parameter that determines the shape of $h(u)$.

\section{Experiments}
\label{sec:exp}
In this section, we first introduce two event-based audio-visual speech recognition datasets used in our experiments and provide the details of the experimental settings. Then, we compare the performance of our proposed HI-AVSNN with existing SNN-based audio-visual fusion methods. Next, we evaluate the noise robustness of our HI-AVSNN and demonstrate its recognition efficiency. Additionally, we investigate the impact of cueing position on performance. Finally, we discuss the energy efficiency of our proposed HI-AVSNN.

\subsection{Datasets}
\paragraph{DVSlip-Audio dataset} This dataset comprises the DVS-Lip dataset along with its corresponding audio files from \cite{tan2022multi}, which simultaneously record the lip movements and speech of volunteers. The lip movements were recorded using the DAVIS346 event camera \cite{brandli2014real} and preprocessed to 128 $\times$ 128 pixels. The audio files were recorded at 44.1kHz and 48kHz due to the variations in recording devices. The dataset consists of 14,896 training samples from 30 volunteers and 4,975 testing samples from the remaining 10 volunteers. The sample durations primarily range from 0.2 to 1.2 seconds, and the dataset covers 100 distinct words. For convenience, we will refer to this event-based audio-visual speech recognition dataset as \textbf{DVSlip-Audio} throughout this paper.
\paragraph{EventLRW dataset} 
This dataset is the event-based version of the LRW dataset \cite{chung2017lip}, a large-scale benchmark for visual speech recognition containing over 500,000 video clips spoken by diverse speakers in real-world settings. Extracted from TV broadcasts, each clip is approximately 1.16 seconds long, consisting of 29 frames. In this paper, we select 100 words from the LRW dataset and convert their corresponding video clips into event streams. We crop the mouth region following the work \cite{feng2021efficient}, and then input the cropped video sequences to the v2e simulator \cite{hu2021v2e}, an event camera simulator, to generate the pixel-level event data. Along with the corresponding audio files, this new event-based audio-visual speech recognition dataset is referred to as \textbf{EventLRW} throughout this paper.

\begin{figure*}[!t]
    \centering
\includegraphics[width=0.8\linewidth]{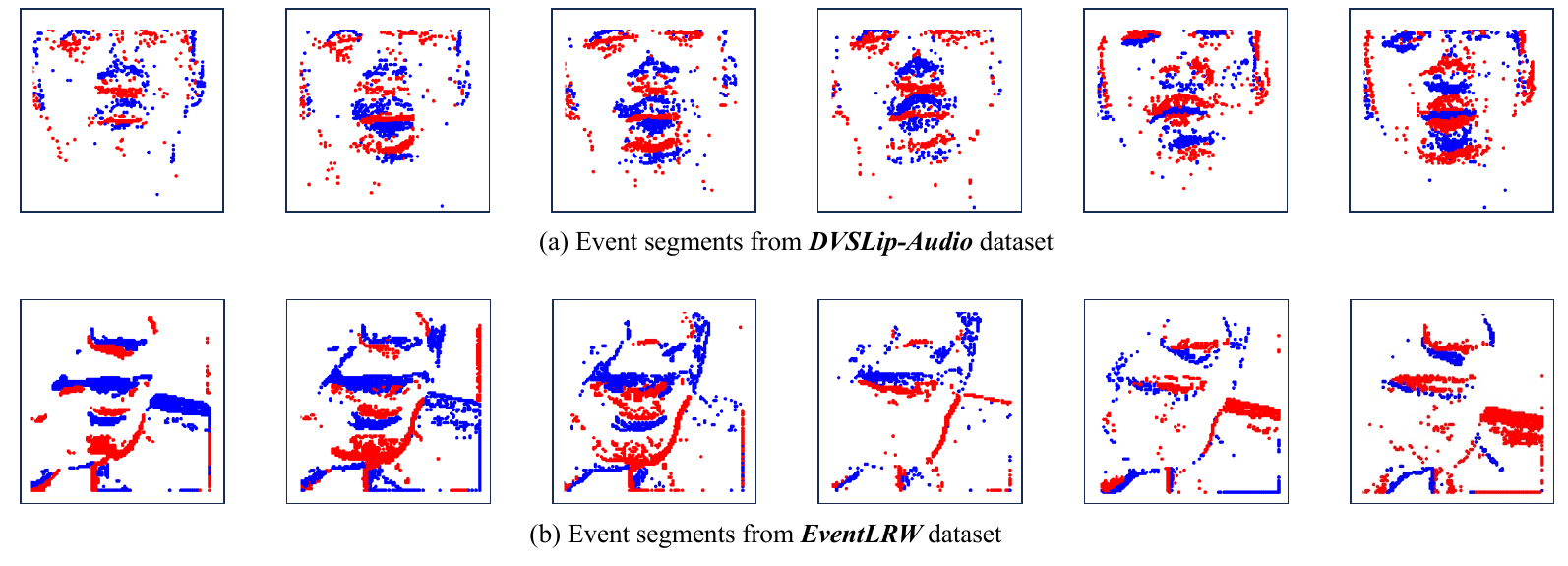}
    \caption{Visualization of samples from two datasets. (a) Event segments from the DVSlip-Audio dataset, captured directly using the DAVIS346 event camera. (b) Event segments from the EventLRW dataset, generated using the v2e simulator to convert video sequences into event streams. Red and blue points represent positive (p=1) and negative (p=-1) events, respectively.}
    \label{fig:visualization}
\end{figure*}

This work marks the first use of both the DVSlip-Audio and EventLRW datasets for AVSR. Previous studies \cite{tan2022multi, bulzomi2023end, liu2024intelligent} have only utilized the visual component of the DVSlip-Audio dataset for lip-reading tasks. The EventLRW dataset is newly created by converting the LRW dataset into event streams, offering a novel dataset for AVSR.  We visualize an example for each dataset in Fig. \ref{fig:visualization}.

\subsection{Implementation Details}
\paragraph{Preprocessing} For visual event streams, we follow the preprocessing pipeline from Tan \textit{et al.} \cite{tan2022multi}. First, central cropping is applied to resize the resolution to $96 \times 96$ pixels. During the training phase, we randomly crop the size to $88 \times 88$ and apply horizontal flipping with a 0.5 probability for data argumentation. For testing, we directly apply the central cropping to $88 \times 88$ pixels. Finally, the spatial resolution of visual events is downsampled to $44 \times 44$ pixels. Event streams from DVSlip-Audio and EventLRW datasets are partitioned into 28 and 29 segments, respectively. For audio data, we first unify the sampling rate of audio files to 44.1kHz through resampling for the DVSlip-Audio dataset. During training, we augment the audio by inverting the polarity with a 0.8 probability, adding a small amount of noise with a 0.1 probability, adjusting the volume with a 0.3 probability, and adding reverb with a 0.6 probability. We then extract 40-dimensional Fbank features using a 120 ms frame size with a 40 ms overlap. The number of frames is standardized to 28 for DVSlip-Audio and 29 for EventLRW to match the number of visual event segments. If the number of frames exceeds the standardized value, we apply linear sampling to reduce it. Otherwise, we pad the frames to the standardized value using zero padding. 

\paragraph{Hyperparameters}
We implement our HI-AVSNN using Pytorch on NVIDIA GeForce RTX 3090 (24GB) GPUs. The SNN is optimized using the Adam optimizer, with a cosine annealing scheduler. The final block in SPN, whether an attention speech block or a speech block, reduces spike activation for decoding. We respectively pre-train the VCE and SPN without sVCSP to initialize the HI-AVSNN. The initial learning rate is 0.001 for pre-training and 0.0005 for fine-tuning. The pre-training phase consists of 150 epochs, while the fine-tuning phase spans 50 epochs. The batch size is 16. The threshold for spike neurons after $QK^TV*s$ is set to 0.5, while for all other neurons, it is set to 1.0. The resting potential is set to 0. The block counts $n_v$, $n_{as}$ and $n_{s1}$, $n_{s}$ are set to 8, 2, 2, and 1, respectively. The initial scaling factor $s$ in the VCSA is set to 0.25, and the constraint parameter $\gamma$ is set to 1.0.

\begin{figure*}
    \centering
    \includegraphics[width=1.0\linewidth]{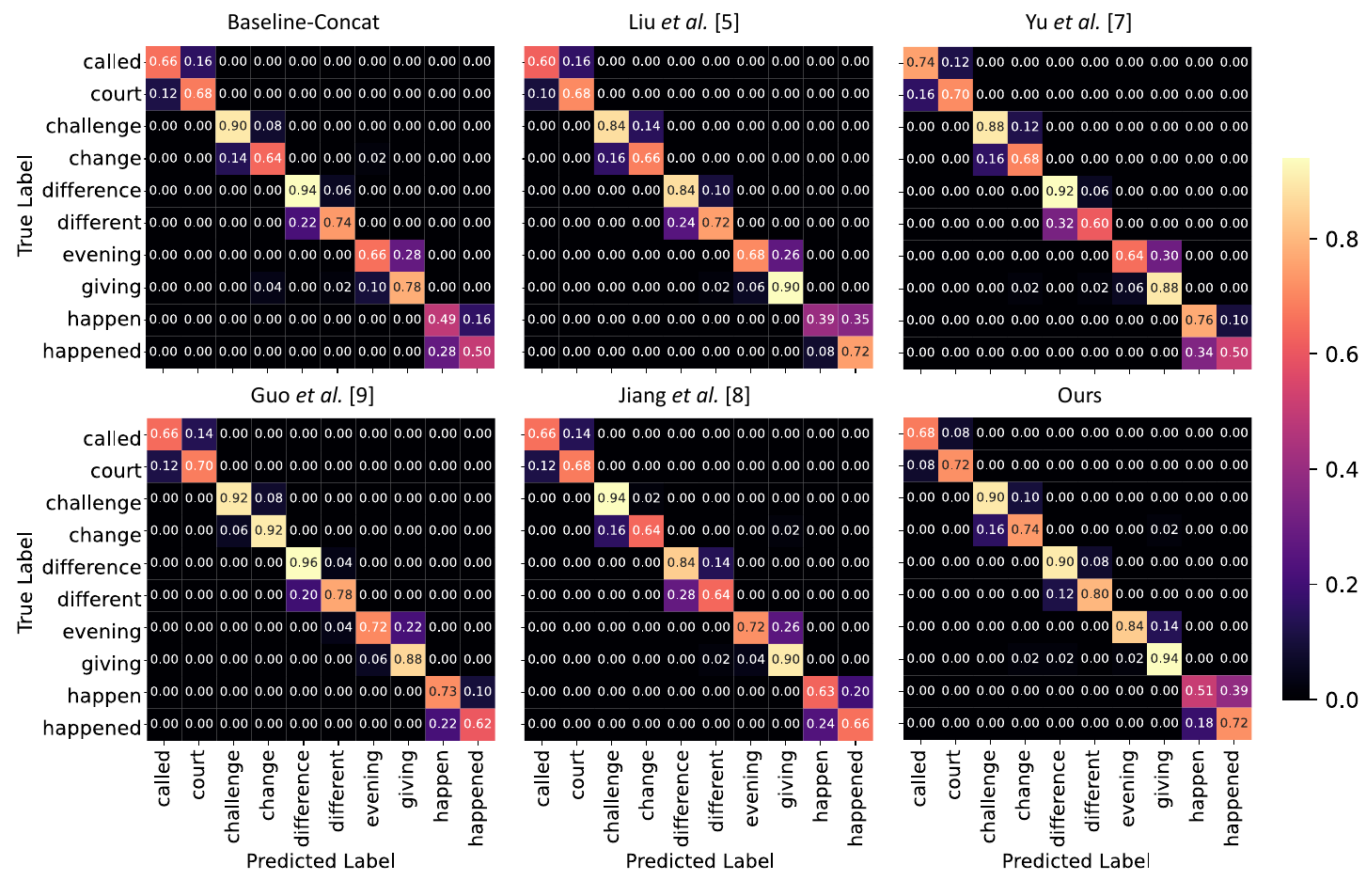}
    \caption{Confusion matrices of 10 words from the DVSlip-Audio dataset. }
    \label{fig:cm}
\end{figure*}
\begin{table*}[!t]
    \begin{center}
    \caption{Performance comparison on the DVSlip-Audio and EventLRW datasets. Acc. represents the accuracy on the entire test set, while Acc.1 and Acc.2 denote accuracy on the first and second parts of the test set, respectively. }
        \resizebox{\linewidth}{!}{
        \begin{tabular}{l|c|c|c|c|c|c|c|c}
            \hline
            \hline
            \multirow{2}{*}{\textbf{Fusion Method}} & \multicolumn{4}{c|}{\textbf{DVSlip-Audio}} & \multicolumn{4}{c}{\textbf{EventLRW}} \\
            \cline{2-9}
            & \textbf{Acc. (\%)} & \textbf{Acc.1 (\%)} & \textbf{Acc.2 (\%)} & \textbf{\#Paras (M)} 
            & \textbf{Acc. (\%)} & \textbf{Acc.1 (\%)} & \textbf{Acc.2 (\%)} & \textbf{\#Params (M)} \\
            \hline
            Baseline-Concat & 83.04 & 91.26 & 74.85 & 10.14 & 94.90 & 95.40 & 94.40 &11.36 \\
            Liu \textit{et al.}\cite{liu2022event} & 84.14 & 92.95 & 75.37 & 10.19 & 96.02 & 97.00 & 95.04 & 11.42 \\
            Yu \textit{et al.} \cite{yu2022multimodal} & 84.26 & 91.42 & 77.14 & 10.14 & 95.86 &  96.16& 95.56 & 11.36 \\
            Guo \textit{et al.} \cite{guo2023transformer} & 84.50 & 91.42 & 77.62 & 19.36 & 94.96 & 96.02& 93.90& 21.85\\
            Jiang \textit{et al.} \cite{jiang2023cmci} & 85.21 & 92.39 & 78.06 & 10.13 & 95.74 & 96.36 & 95.12 & 11.35\\
            \textbf{Ours} & 86.41 & 93.15 & 79.70 & 10.57 & 96.76 & 97.40 &  96.12 & 13.03 \\
            \hline
            \hline
        \end{tabular}}
        \label{table:performance}
    \end{center}
\end{table*}

\paragraph{Comparison Methods}
We evaluate five SNN-based audio-visual methods for comparison, ensuring fairness by adjusting their networks to maintain a similar number of parameters as our HI-AVSNN and applying the same visual and audio frontends. The first method serves as our baseline, which fuses audio and visual features by concatenating them at a high-level layer (before the classification layer). This straightforward fusion approach is widely used as a baseline in related works \cite{liu2022event,jiang2023cmci}.
The second and third methods are from Liu \textit{et al.} \cite{liu2022event} and Jiang \textit{et al.} \cite{jiang2023cmci}, with one designed for digit recognition and the other for object recognition.
The fourth method is from Yu \textit{et al.} \cite{yu2022multimodal}, which is the only existing SNN designed for audio-visual speech recognition. However, it operates non-causally. To ensure fairness, we adapt it to our causal framework for evaluation.
The fifth method is from Guo \textit{et al}. \cite{guo2023transformer}, which is also a non-causal SNN. Since it processes speech spectrograms as single images and applies convolution operations directly on these images, modifying it to integrate into our causal framework would require significant changes to its original design. Thus, we evaluate this method in its original form.

\subsection{Performance Comparison}
Table \ref{table:performance}\footnote{The results of comparing SNNs are based on our own implementation and optimization, as there is no publicly available code.} presents a comprehensive comparison between our proposed HI-AVSNN and other state-of-the-art SNN-based audio-visual multimodal fusion methods. Our SNN obtains the classification accuracy of 86.41\% and 96.76\% on the DVSlip-Audio and EventLRW datasets, respectively, surpassing all other methods. Notably, on the DVSlip-Audio dataset, the fusion method from the only existing SNN-based AVSR \cite{yu2022multimodal} achieves an accuracy of 84.26\%, which is over 2\% lower than ours. 
The method from \cite{guo2023transformer}, despite having nearly twice the parameters and processing data non-causally, also delivers lower accuracy compared to ours. These results highlight the computational advantages of our human-inspired approach to fusing speech and visual information. Additionally, we observe that methods consistently achieve higher accuracy on the EventLRW dataset than on DVSlip-Audio. This result can be attributed to two reasons: (1) DVSlip-Audio contains low-quality audio files with noise and variations introduced by different recording devices, which degrade recognition performance; and (2) DVSlip-Audio has over 6 times fewer samples than EventLRW, which limits the model’s ability to learn effectively and generalize across different conditions. Despite these challenges, our HI-AVSNN still achieves the highest accuracy on DVSlip-Audio, demonstrating its robustness and adaptability across varying data conditions.

Furthermore, we divided the words in each dataset into two parts: the first part consists of 50 distinct words, and the second part comprises another 25 pairs of visually or phonetically similar words. Table~\ref{table:performance} reports the corresponding accuracies as Acc.1 and Acc.2. Across all methods, Acc.1 values are consistently higher than Acc.2, reflecting the increased difficulty of distinguishing words that are visually or phonetically similar. Nevertheless, our HI-AVSNN delivers superior performance in both parts across both datasets. In particular, it outperforms other methods by nearly 5\% on the second part of DVSlip-Audio. The confusion matrices of 10 words from this second part are shown in Fig. \ref{fig:cm}. Words with similar pronunciation or lip movements are placed adjacent to each other. We can see that our method exhibits less confusion; for example, it misclassifies "different" as "difference" 12\% of the time, whereas this error rate increases to 20\%--32\% with other methods. This benefit arises because our method does not merely concatenate or add features from the two modalities but selectively utilizes the most relevant information from each modality. These results further emphasize the strengths of our human-inspired computing approach in enabling robust and efficient information processing. 

\subsection{Noise Robustness}
\begin{table*}[!t]
\centering
  \tabcolsep=0.5cm
  \caption{Performance comparison in noisy environments. Samples in the DVSlip-Audio dataset that naturally contain noise are labeled as 'noise*'.}
 \resizebox{1.0\linewidth}{!}{
\renewcommand{\arraystretch}{1.2}
\begin{tabular}{|c|l|c|c|c|c|c|c|c|}
\cline{3-9}
\multicolumn{2}{c|}{} & \multicolumn{6}{c|}{SNR (dB)} & \\
\cline{3-9}
\multicolumn{2}{c|}{} & Clean & Noise* & 10 & 5 & 0 & -5 & Average \\
\cline{3-9}
\multicolumn{2}{c|}{} & \multicolumn{7}{c|}{\textbf{DVSlip-Audio}} \\
\hline
\multirow{4}{*}{Method} & Vision-only & \multicolumn{7}{c|}{50.03} \\
\cline{2-9}
& Audio-only & 86.23 & 70.90 & 82.77 & 81.83 & 76.00 & 62.43 & 77.05 \\
\cline{2-9}
& Baseline-concat & 86.40 & 75.39 & 84.30 & 83.03 & 80.10 & 73.87 & 80.83\\
\cline{2-9}
& Ours & 87.60 & 80.05 & 85.77 & 85.27 & 82.17 & 76.87 & 83.13 \\
\hline
\multicolumn{2}{c|}{} & \multicolumn{7}{c|}
{\textbf{EventLRW}} \\
\hline
\multirow{4}{*}{Method} & Vision-only & \multicolumn{7}{c|}{63.70} \\
\cline{2-9}
 & Audio-only  & 	94.98 & -& 94.24	& 93.12 & 	89.88& 	81.52 & 90.75 \\
\cline{2-9}
& Baseline-concat & 94.72 & - &	94.12 & 93.60 & 91.78 & 87.00 & 92.24\\
\cline{2-9}
& Ours & 96.36 & - & 95.74 & 95.64 & 93.84  & 91.10 & 94.54 \\
\hline
\end{tabular}
\renewcommand{\arraystretch}{1.0}
}
\label{table:noise}
\end{table*}
In this section, we verify the noise robustness of our proposed HI-AVSNN. Due to the recording device and environments, some of the original audio files in the DVSlip-Audio dataset naturally contain noise. We categorize these files as ``noise*", while the original files without noise are labeled as ``clean". During training, we add babble noise at signal-to-noise ratio (SNR) levels of 10 dB, 5 dB, 0 dB, and -5 dB to the clean files. The babble noise is generated by mixing samples as presented in \cite{afouras2018deep}. The trained SNN models are then tested on these different noise levels to evaluate their robustness.

As shown in Table \ref{table:noise}, our proposed HI-AVSNN achieves the highest accuracy in both clean and noisy environments across both datasets, demonstrating its robustness and effectiveness. Comparing the audio-only method to audio-visual multimodal solutions, we observe that multimodal integration significantly reduces the impact of noise on recognition accuracy.  For example, on the DVSlip-Audio dataset, the performance of the audio-only method drops significantly as noise levels increase, falling from 86.23\% in the clean condition to 62.43\% at -5 dB, a decline of over 23\%. In contrast, audio-visual multimodal solutions, including the baseline and our HI-AVSNN, experience a decline of less than 11\%, indicating the advantage of integrating visual information when audio quality is degraded by noise. Moreover, our HI-AVSNN consistently outperforms the baseline method, particularly at higher noise levels. Across both datasets, our HI-AVSNN surpasses the baseline method by 1\% in the clean condition and extends this margin to over 3\% in SNR -5 dB conditions. This demonstrates the effectiveness of our human-inspired fusion strategy in enhancing noise robustness, ensuring more reliable performance under challenging conditions.

\subsection{Recognition Efficiency}
 \begin{figure}
     \centering
    \includegraphics[width=0.95\linewidth]{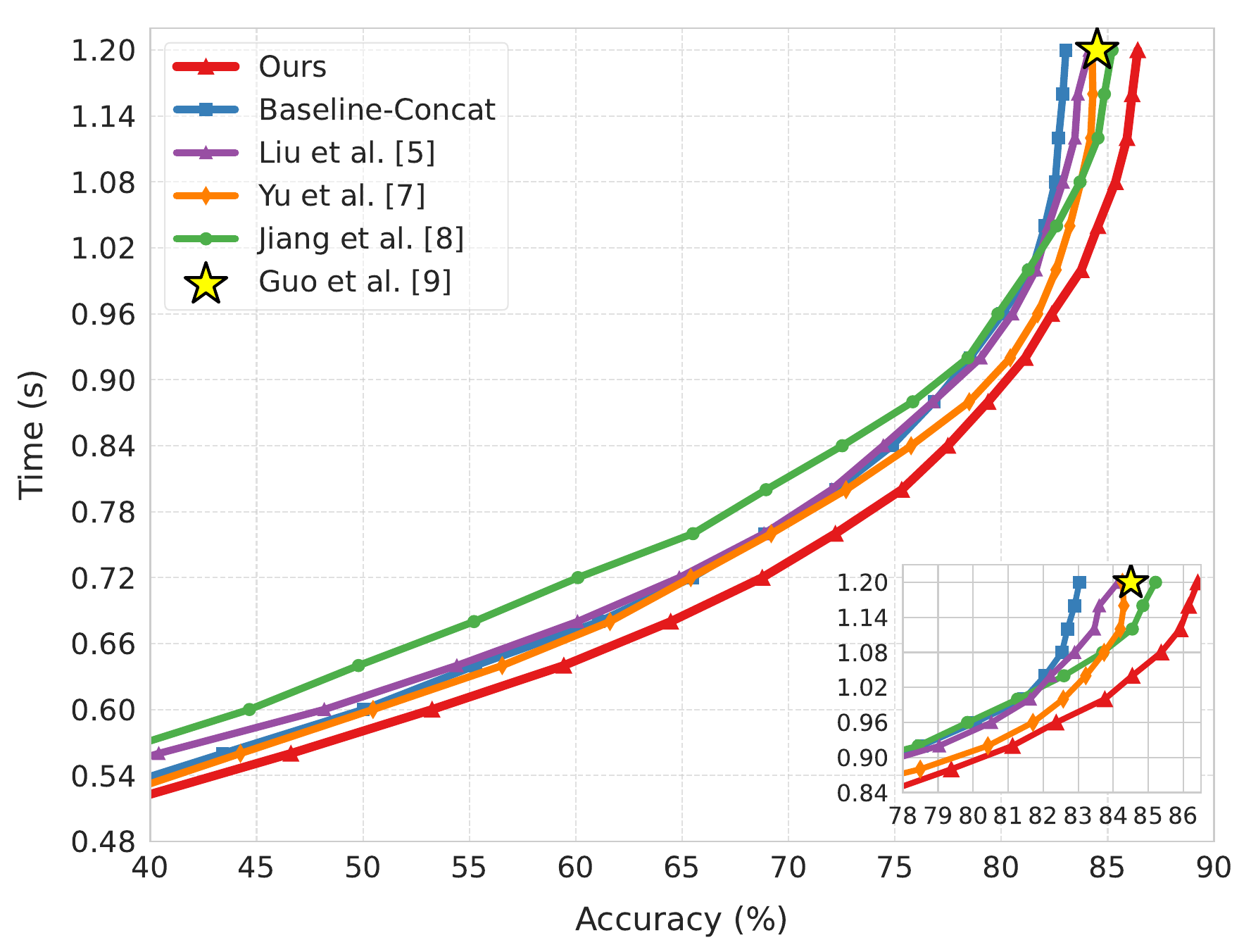}
     \caption{Compare the time used to achieve the same accuracy on the DVSlip-Audio dataset. A shorter time indicates higher recognition efficiency, meaning the model can achieve the same accuracy with less processed input. }
     \label{fig:sequence}
 \end{figure}
In this section, we assess the recognition efficiency of our proposed HI-AVSNN by comparing the time required to reach a given accuracy against existing audio-visual multimodal SNNs. Here, time refers to the duration of the input processed by the network. We begin the evaluation from half of the final accuracy as the available information is extremely limited at the very beginning, making the accuracy too low for a meaningful comparison. Some existing methods \cite{yu2022multimodal,guo2023transformer} operate non-causally and cannot provide accuracy at each time length. To facilitate comparison, we adapt their fusion methods to our causal framework. However, Guo \textit{et al.} \cite{guo2023transformer} treats speech spectrograms as single images and applies convolution operations directly, requiring substantial modifications for causal integration. Thus we only report its final accuracy.

As shown in Fig. \ref{fig:sequence}, recognition accuracy improves across all methods as more input information becomes available. Our HI-AVSNN consistently outperforms all other SNNs, achieving the same accuracy in a shorter time. This efficiency advantage is particularly evident in the later stages, where additional accuracy improvements require processing more input. For example, to reach an accuracy of 83\%, HI-AVSNN requires 0.98s, while the baseline method, Liu \textit{et al.} \cite{liu2022event} and Jiang \textit{et al.} \cite{jiang2023cmci} take approximately 22\%, 10\% and 7\% more time compared to ours, respectively, highlighting the superior ability of our HI-AVSNN to quickly recognize speech content, underscoring its exceptional recognition efficiency in audio-visual speech recognition. \cite{guo2023transformer} can only produce results after receiving the entire input sequence due to its reliance on future information. Despite this, our HI-AVSNN still surpasses it in accuracy, further demonstrating the efficiency and flexibility of our human-inspired computing approach.

\subsection{Position of Cueing in sVCSP}
    \begin{figure}
     \centering
    \includegraphics[width=0.95\linewidth]{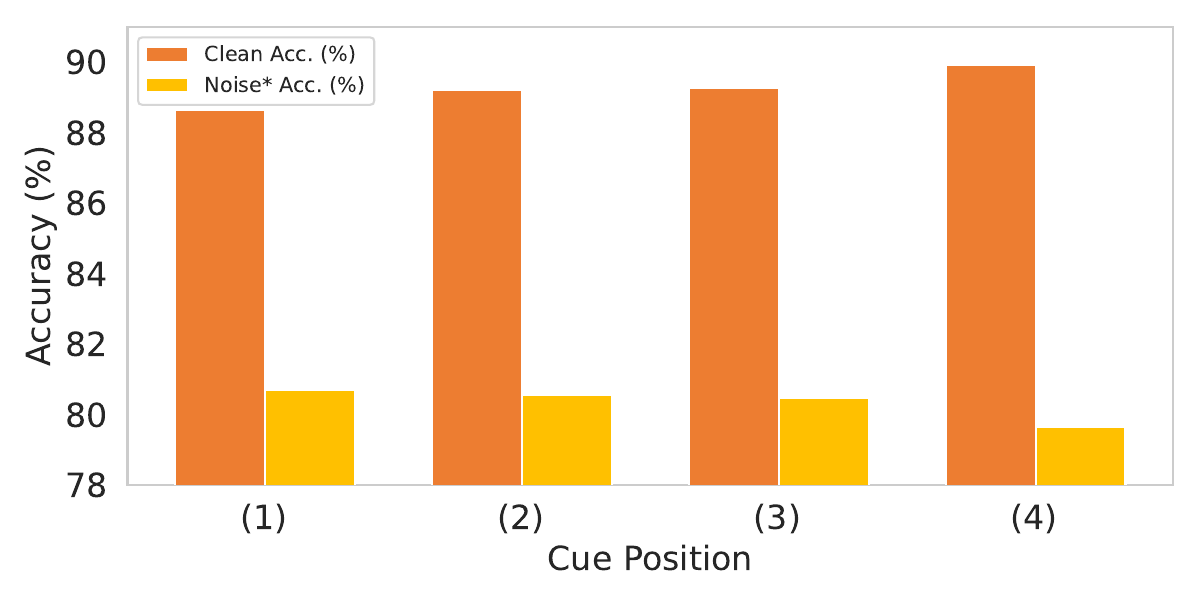}
     \caption{Ablation study of the position of single cueing. Clean Acc. and Noise* Acc. means the accuracy on clean samples and noise samples from the original DVSlip-Audio dataset.}
     \label{fig:ab1}
 \end{figure}
   \begin{figure}
     \centering
    \includegraphics[width=0.95\linewidth]{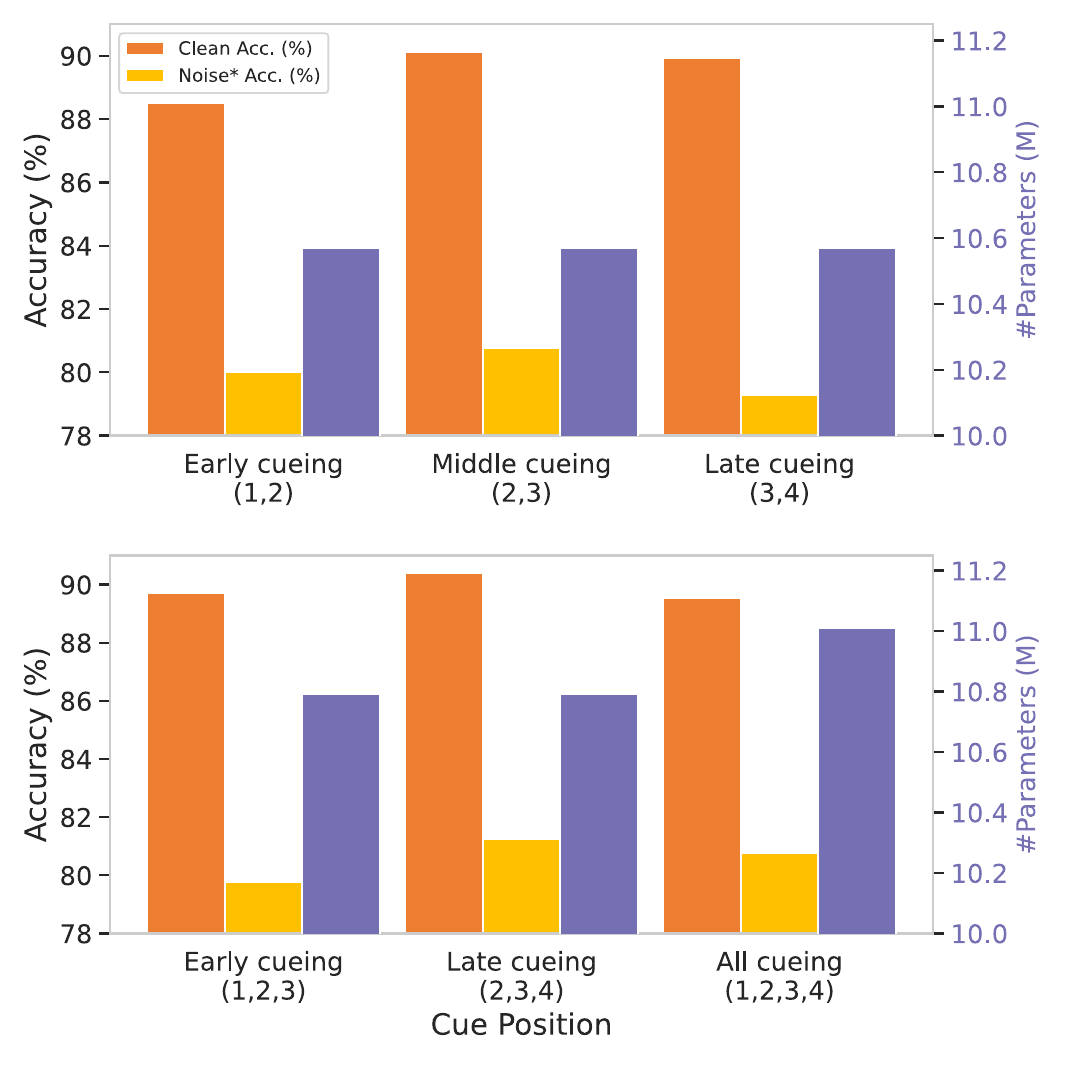}
     \caption{Ablation study of the positions and counts of multiple cueing.}
     \label{fig:ab2}
 \end{figure} 
In this section, we investigate the impact of visual cueing position on recognition performance. We conduct two groups of ablation experiments: one focusing on the position of single cueing and the other on multiple cueing positions. 
To ensure consistency with comparison methods, we structure the SPN with four blocks, each serving as a candidate for visual cueing, referred to as (1), (2), (3), and (4).  If a block receives visual cues, it functions as an attention speech block; otherwise, it remains a speech block. The experiments are conducted on the DVSlip-Audio dataset.

For single cueing, we apply VCSA at each of the four positions separately and compare the recognition accuracy on clean and noisy audio samples from the DVSlip-Audio dataset. As shown in Fig. \ref{fig:ab1}, accuracy on clean samples increases from 88.63\% to 89.93\% as the cueing position shifts from the first block to the fourth block, whereas accuracy on noisy samples declines from 80.71\% to 79.65\%. This trend arises because late cueing, which occurs closer to the final decision-making layers, enhances clean accuracy by refining well-formed speech representations with visual cues. In contrast, early cueing helps the network focus on relevant speech features from the start, enhancing robustness to noise interference. These findings highlight a trade-off: while late cueing optimally refines high-quality speech representations, early cueing provides crucial guidance for robust speech recognition in noisy environments.

Next, we explore multiple cueing to combine the advantages of different cueing positions. As shown in Fig. \ref{fig:ab2}, we first examine two-position cueing configurations: (1,2), (2,3), and (3,4). Similar to single cueing, early cueing (1,2) results in lower accuracy on clean samples, while late cueing (3,4) leads to lower accuracy in noisy conditions. However, a notable difference is that the middle cueing configuration (2,3) achieves the highest accuracy on both clean and noisy samples. This superior performance is attributed to its optimal balance between early guidance and late refinement, leveraging the strengths of both early cueing and late cueing. We then evaluate three-position cueing configurations, (1,2,3) for early cueing and (2,3,4) for late cueing. We observe that late cueing outperforms early cueing on clean samples, which aligns with the findings from single cueing. On the noise samples, late cueing also surpasses early cueing. These results suggest that while early cues provide advantages, using multiple early cues offers limited improvement in accuracy. This is likely because speech features are not yet sufficiently extracted at early stages, making them less representative for fusion with visual cues compared to later cueing. We finally employ an SNN with cueing at all four positions (1,2,3,4) and found that it underperforms (2,3,4) on both kinds of samples, indicating that more cueing positions do not necessarily lead to higher accuracy. To conclude, for multiple cueing, mid-to-late layer integration (2,3,4) or (2,3) proves to be the most effective strategy. Since (2,3) achieves comparable accuracy while requiring fewer parameters, we adopt (2,3) as the default cueing configuration throughout this paper unless otherwise specified.

 \begin{table*}[!t]
    \begin{center}
    \caption{Energy cost comparison for a single forward.}
        \resizebox{\linewidth}{!}{
        \begin{tabular}{l|c|c|c|c|c|c|c|c}
            \hline
            \hline
            \multirow{2}{*}{\textbf{Fusion Method}} & \multicolumn{4}{c|}{\textbf{DVSlip-Audio}} & \multicolumn{4}{c}{\textbf{EventLRW}} \\
            \cline{2-9}
            & \textbf{Acc. (\%)} & \textbf{\#Mul. (M)} & \textbf{\#Add. (M)} & \textbf{Energy (mJ)} 
            & \textbf{Acc. (\%)} & \textbf{\#Mul. (M)} & \textbf{\#Add. (M)} & \textbf{Energy (mJ)} \\
            \hline
            ANN (ResNet+LSTM) & 87.34 & 455.6 & 455.6 & 2.10 & 97.06 & 640.9 & 640.9 & 2.95\\
            Baseline-Concat & 83.04 & 31.5 & 1048.3 & 1.06  & 94.90 & 33.8 & 947.5& 0.98\\
            Liu \textit{et al.} \cite{liu2022event} & 84.14 & 31.5  & 1143.6  & 1.15  & 96.02 & 33.8& 769.5 & 0.82 \\
            Yu \textit{et al.} \cite{yu2022multimodal} & 84.26 & 31.5 & 1116.0 & 1.12 & 95.86 & 33.9 & 937.6 & 0.97\\
            Guo \textit{et al.} \cite{guo2023transformer} & 84.50 & 49.4 & 2753.0 & 2.66  & 94.96 & 52.2 &2787.8 & 2.70\\
            Jiang \textit{et al.} \cite{jiang2023cmci} & 85.21 & 31.5 & 1105.7 & 1.11 & 95.74 & 33.8 & 1194.9 & 1.20 \\
            \textbf{Ours} & 86.41 & 32.8  & 1200.0 & 1.20 & 96.76 & 34.7 &  1024.6 & 1.05 \\
            \hline
            \hline
        \end{tabular}}
        \label{table:energy}
    \end{center}
\end{table*}

 \subsection{Energy Consumption}
 In this section, we estimate the theoretical energy consumption of our HI-AVSNN and compare it with other works using the common approach in the neuromorphic computing community \cite{zhouspikformer, yao2024spike, zhang2025spiking}. 
The energy consumption is calculated based on 45nm CMOS technology, where an addition operation requires 0.9 pJ energy and a multiplication operation consumes 3.7 pJ \cite{horowitz20141}. For ANNs, the operation count per layer is calculated as follows:
\begin{equation}
    \#OP_{ANN}^{conv} = k_w*k_h*c_{in}*c_{out}*h_{out}*w_{out}
\end{equation}
\begin{equation}
    \#OP_{ANN}^{fc} = f_{in}*f_{out}
\end{equation}
where $k_w$ and $k_h$ denote the width and height of the convolutional kernel, respectively, $c_{in}$ and $c_{out}$ represent the input and output channels, $h_{out}$ and $w_{out}$ refer to the height and width of the output feature map, and $f_{in}$ and $f_{out}$ denote the number of input and output features.
For SNNs, the computational cost depends on both the network architecture and the spike rate. The operation count in each SNN layer can be expressed as:
\begin{equation}
   \#OP_{SNN} = \sum_t^T SpikeRate_t * \#OP_{ANN}
\end{equation}
\begin{equation}
    SpikeRate_t = \frac{\#TotalSpikes_t}{\#Neurons}
\end{equation}
where $\#TotalSpikes_t$ and $SpikeRate_t$ are
the total spikes and the average spike rate at the $t$-th timestep of the current layer, respectively. 

Table \ref{table:energy} shows the energy consumption of our HI-AVSNN alongside other comparison networks. We also include an ANN-based model incorporating ResNet for visual processing and LSTM for audio processing. Our HI-AVSNN consumes approximately 1.20 mJ with an accuracy of 86.41\% on the DVSlip-Audio dataset and 1.05 mJ with an accuracy of 96.76\% on the EventLRW dataset. Compared to other SNN methods, HI-AVSNN achieves higher accuracy while using half the energy of Guo \textit{et al.} \cite{guo2023transformer}, demonstrating its energy efficiency. Additionally, although our SNN consumes  0.05 mJ and 0.08 mJ more energy than Liu \textit{et al.} \cite{liu2022event} and Yu \textit{et al.} \cite{yu2022multimodal} on the DVSlip-Audio dataset, it outperforms them by over 2\% in accuracy. Meanwhile, the ANN-based model consumes 0.9 mJ more energy than ours but achieves less than a 1\% accuracy improvement. These results highlight the effectiveness of our method in balancing energy efficiency and recognition performance, making it a more practical choice for real-world applications. We acknowledge that energy consumption arises not only from computation but also from memory access, which involves hardware design considerations beyond the scope of our study. Nevertheless, it's worth noting that the binary nature and greater sparsity of our HI-AVSNN can help reduce access-related energy consumption.

\section{Conclusion}
\label{sec:conclusion}
This paper proposes a human-inspired audio-visual speech recognition SNN, incorporating three key computing characteristics of human speech perception: (1) spike activity, (2) cueing interaction, and (3) causal processing. For spike activity, we utilize an SNN architecture to process information, complemented by an event camera for efficient capture of lip movements. For cueing interaction, we introduce the sVCSP scheme, where visual features incrementally guide the auditory processing to focus on critical auditory features.  For causal processing, we align the SNN's temporal dimension with visual and auditory features and enforce temporal masking to ensure that only past and current information is utilized, enabling real-time, low-latency operation. Experimental results on the DVSLip-Audio and EventLRW datasets highlight the superior performance of HI-AVSNN compared to other audio-visual multimodal fusion methods, demonstrating the effectiveness of our human-inspired computing paradigms. Additionally, our HI-AVSNN exhibits robust noise tolerance and efficient recognition, making it highly suitable for real-world applications. Ablation studies further quantify the impact of cueing interaction, while energy consumption analysis underscores the system’s energy efficiency. Our work marks a step forward in brain-inspired computing, offering a highly efficient and robust solution for real-world audio-visual speech recognition tasks.

\bibliography{main}
\bibliographystyle{IEEEtran}
\end{document}